\pdfoutput=1 \documentclass[aps,prd,amsmath,floats,floatfix, twocolumn,superscriptaddress,nofootinbib,showpacs,longbibliography]{revtex4-2}

\usepackage{amsmath}
\usepackage[T1]{fontenc}
\usepackage[utf8]{inputenc}
\usepackage{lmodern}
\usepackage{cancel}
\usepackage{verbatim}

\usepackage[dvipsnames, usenames]{xcolor}
\definecolor{linkcolor}{rgb}{0.0,0.3,0.5}
\usepackage[hypertexnames=false, unicode, colorlinks=true, linkcolor=linkcolor,citecolor=linkcolor, filecolor=linkcolor, urlcolor=linkcolor, pdfusetitle]{hyperref}

\usepackage[all]{hypcap}
\usepackage{graphicx}
\usepackage{xspace}
\usepackage{amssymb}
\usepackage[normalem]{ulem} 
\usepackage{bm}
\usepackage{float}

\usepackage[caption=false]{subfig}

\usepackage{microtype}

\usepackage[english]{babel}
\usepackage{blindtext}

\usepackage{etoolbox} 
\usepackage{orcidlink}

\graphicspath{%
  {figs/}%
}

\DeclareMathAlphabet{\mathpzc}{OT1}{pzc}{m}{it}

\begin{document}

\title{On relativistic observables in black bounce spacetimes}

\newcommand{\Cornell}{\affiliation{Departamento de Física Fundamental, Universidad de Salamanca, 37008 Salamanca, Spain}}
\newcommand{\Caltech}{\affiliation{Cosmology and Gravity Group, Department of Mathematics and Applied Mathematics, University of Cape Town, Rondebosch 7700, Cape Town, South Africa}}

\author{Víctor Ovejero-Bermúdez}
\email{idu064397@usal.es}
\Cornell

\author{Álvaro de la Cruz-Dombriz}
\email{alvaro.dombriz@usal.es}
\Cornell\Caltech

\author{Riccardo Della Monica}
\email{rdellamonica@tecnico.ulisboa.pt}
\affiliation{CENTRA, Departamento de Física, Instituto Superior Técnico – IST\\
Universidade de Lisboa – UL, Avenida Rovisco Pais 1, 1049-001 Lisboa, Portugal}

\pacs{04.50.Kd, 98.80.-k, 98.80.Cq, 12.60.-i}

\hypersetup{pdfauthor={Habib et al.}}

\begin{abstract}
We investigate the phenomenology of black bounce spacetimes through a combined analytical and numerical study of relativistic observables associated with both time-like and null geodesics. Black bounce geometries provide a continuous interpolation between Schwarzschild black holes and traversable wormholes by introducing a regularization bounce parameter $\alpha$, which removes the central singularity by replacing it with a finite-radius throat. Using the Hamilton--Jacobi and Lagrangian formalisms, we derive weak-field analytical expressions for the periastron advance of massive particles and the deflection angle of light, highlighting the leading corrections induced by the bounce parameter. These results are systematically compared with numerical integrations performed with the \texttt{PyGRO} code. We show that increasing $\alpha$ enhances both the periastron precession and the light deflection with respect to the Schwarzschild case. Additionally, we analyze the relativistic redshift of bound orbits and identify cumulative temporal phase shifts associated with the modified geometry. For null geodesics, we also determine numerically the critical impact parameter separating scattered and captured (or transmitted through the throat) photon trajectories. Although this quantity remains identical to the Schwarzschild value in the regular black hole branch, it increases significantly in the traversable wormhole regime, providing a clear observational signature of the throat structure. Our results demonstrate that black bounce spacetimes can produce measurable deviations from standard black hole predictions, opening the possibility of constraining these geometries with current and future observations of strong gravitational fields.
\end{abstract}

\maketitle

\section{Introduction}

General Relativity (GR) remains the fundamental theoretical framework for the description of gravitation, successfully accounting for phenomena across an extraordinary range of physical scales. From classical Solar-System tests \cite{will2014confrontation} to the recent direct detections of gravitational waves by the LIGO--Virgo--KAGRA collaboration~\cite{abbott2016observation} and the direct imaging of supermassive compact objects by the Event Horizon Telescope (EHT)~\cite{
EventHorizonTelescope:2019dse, EventHorizonTelescope:2022wkp}, 
the predictions of GR have been confirmed with remarkable precision. In para\-llel, the high-accuracy monitoring of the S-stars cluster around Sagittarius A* (Sgr A*) by the  GRAVITY collaboration has provided unprecedented access to the strong-field regime of gravity through the measurements of both the relativistic redshift and the Schwarzschild precession of the S2 star~\cite{abuter2018redshift,abuter2020precession}, which represents a milestone in verifying the classical tests of GR around supermassive compact objects.
These developments have opened a new era in observational gravity, where the structure of compact objects can be probed with increasing precision.

Despite its extraordinary success, GR predicts the exis\-tence of spacetime singularities under rather generic conditions~\cite{hawking1970singularities}. In black hole spacetimes, curvature invariants diverge at the center of the geometry, signaling the breakdown of the classical description of spacetime. This issue has motivated the exploration of regular black holes and black hole mimickers, namely compact geometries capable of reproducing the successful predictions of GR at astrophysical scales while avoiding singular behavior in the strong-field region~\cite{cardoso2019testing}. Such alternatives include gravastars~\cite{pani2015love, mottola2023gravitational}, boson stars~\cite{liebling2023dynamical, Visinelli2021}, regular black holes~\cite{ansoldi2008spherical, lan2023regular}, and traversable wormholes~\cite{garcia2012generic, simpson2023black}, among many others.

In this context, black bounce geometries have recently attracted considerable attention. Originally introduced by Simpson and Visser~\cite{simpson2019black}, these spacetimes provide a continuous interpolation between Schwarzschild black holes and traversable wormholes through the introduction of a regularization parameter $\alpha$. Instead of a central singularity, the geometry contains a finite-radius throat that regularizes the spacetime curvature. Depending on the value of $\alpha$, the geometry can describe a regular black hole, a one-way wormhole, or a two-way traversable wormhole. This unified framework offers a particularly convenient setting for investigating how modifications of the near-horizon structure affect relativistic observables while preserving the standard Schwarzschild behavior at large distances.

The growing observational precision achieved in recent years, both across the electromagnetic spectrum and through gravitational waves, has stimulated extensive interest in identifying measurable signatures of these regularized compact objects. Their observational and phenomenological properties have been explored in se\-ve\-ral directions, including weak- and strong-field gravitational lensing~\cite{Nascimento:2020ime,Tsukamoto:2020bjm}, gravitational-wave echoes and quasi-normal ringing~\cite{Yang:2021cvh,Santos:2025xbk,Yang:2026jch}, the optical appearance and radiative properties of thin accretion disks~\cite{Guerrero:2021ues, Bambhaniya:2021ugr}, and tidal effects along geodesic congruences~\cite{Arora:2023ltv}. Further developments have considered generalized black bounce metrics and their shadow structure~\cite{Nascimento:2025mtr}, possible realizations in modified-gravity or braneworld settings~\cite{Ling:2025ncw,Crispim:2024yjz}, matter sources for black bounce geometries~\cite{Lessa:2024erf}, and the structure of periodic orbits~\cite{Zhang:2022zox, Bragado:2025jrg}. Moreover, the black bounce model has been used to put constraints on the fundamental nature of the supermassive compact object at the center of the Milky Way \cite{DellaMonica:2021fdr}. These results show that the bounce parameter can affect a broad class of ob\-ser\-va\-bles, motivating a systematic comparison between weak-field analytical predictions and fully numerical geodesic integrations.

These observational signatures are ultimately encoded in the motion of matter and radiation propagating through spacetime geometry. For instance, the motion of stars orbiting compact objects provides direct information about the underlying gravitational potential through ob\-ser\-va\-bles such as the periastron advance and the relativistic redshift. Likewise, null geodesics determine gravitational lensing phenomena, photon spheres, and shadow pro\-per\-ties. Since these observables depend directly on the spacetime metric, they constitute a particularly sensitive framework for distinguishing between Schwarzschild black holes and regularized geometries. Moreover, even small deviations from the Schwarzschild metric can accumulate over multiple orbital periods, potentially producing observable signatures accessible to current and future astrophysical facilities. 
As highlighted in recent lite\-ra\-ture~\cite{Zhang:2022zox}, different physical attributes \textemdash such as the presence of a classical electric charge, the spin of the central body or a topological regularization \textemdash can produce similar cumulative effects on orbital motion, light deflection or frequency modulation. For instance, the detected relativistic precession and redshift profiles of the S2 star \cite{abuter2018redshift, abuter2020precession} could, in principle, be reproduced by a variety of models within current observational uncertainties. Breaking this degeneracy requires a systematic and comparative analysis of multiple observables. It thus is essential not only to propose new geometries but to identify unique ``signatures'' that can distinguish between fundamentally different physical mechanisms. The combined study of spatial observables (precession and deflection) and temporal ones (redshift) is crucial for this purpose.

In this work, we investigate the phenomenology of the Simpson--Visser black bounce spacetime through a combined analytical and numerical study of time-like and null geodesics. We derive weak-field analytical expressions for the periastron precession of massive particles and for the deflection angle of light, identifying the leading corrections induced by the bounce parameter $\alpha$. These analytical predictions are systematically compared against high-precision numerical integrations performed with the \texttt{PyGRO} code~\cite{della2025pygro}. Additionally, we analyze the relativistic redshift associated with bound time-like orbits and numerically determine the critical impact parameter governing photon capture and transmission in the wormhole regime.

This paper is organized as follows. In Sec. \ref{sec:theory}, we introduce the theoretical framework of the black bounce spacetime and discuss its geometrical and causal properties in both Simpson--Visser and areal coordinates. In Sec.~\ref{sec:timelikegeo}, we study time-like geodesics using the Hamilton--Jacobi formalism and derive an analytical expression for the periastron advance in the weak-field limit. We then compare the analytical predictions with numerical integrations and analyze the corresponding relativistic redshift. Subsequently, in Sec.~\ref{sec:nullgeo}, we focus on null geodesics through a Lagrangian approach, deriving the weak-field deflection angle and comparing it against numerical simulations. We also determine the photon-sphere radius and the critical impact parameter that separates scattered and captured (or transmitted) photon trajectories. Finally, in Sec.\ref{sec:Conclusions}, we summarize our main conclusions and discuss the observational implications of black bounce geometries for current and future strong-gravity experiments.
Throughout this work we use geometrized  units $c=G=1$ and adopt the metric signature  $\{-, +, +, +\}$. Furthermore, all physical quantities are expressed in terms of the black bounce mass $M$. In our numerical simulations and figures, we adopt the convention $M = 1$. Consequently, the results are presented in dimensionless units and can be rescaled to any physical mass. Finally, in what follows, unless otherwise specified, we have resorted to resolve exact equations numerically with the open-source code \texttt{PyGRO} \cite{della2025pygro}.

\section{Theoretical Framework}
\label{sec:theory}

The determination of the vacuum spacetime geometry ($T_{\mu\nu} = 0$) in the presence of a spherical mass and static source begins by considering a general ansatz for a spherically symmetric metric
\begin{equation}\label{metsph}
    {\rm d}s^2 = -A(r){\rm d}t^2 + B(r){\rm d}r^2 + C(r)({\rm d}\theta^2+\sin^2\theta\, {\rm d}\varphi^2),
\end{equation}
where for the angular sector the usual spherical coordinates ($\theta$, $\varphi$) have been employed and the metric coefficients are expressed solely as functions of a generic radial coordinate $r$.
Upon introducing this metric form into the GR vacuum field equations, it is possible to uniquely solve for the unknown functions that compose the metric above. Specifically, in pure vacuum the Schwarzschild spacetime solution 
\begin{align}\label{schwmetric}
    {\rm d}s^2 = & -\left(1-\frac{2M}{r}\right){\rm d}t^2 + \left(1-\frac{2M}{r}\right)^{-1}{\rm d}r^2 \nonumber\\ & + r^2({\rm d}\theta^2+\sin^2\theta\,{\rm d}\varphi^2)\,,
\end{align}
is obtained, being characterized by one single parameter: the mass $M$ of the gravitational source.
Despite its fundamental role in describing black holes, the singular behavior of the Schwarzschild geometry at the origin
($r=0$) has motivated the exploration of alternative geometries, namely classes of exotic compact objects that, while remaining within the framework of GR, i.e. without considering extensions or modifications to the underlying theory of gravity, could mimic black hole behavior at large and intermediate distances, while presenting properties in the strong-field regime (e.g. a regularization of the central singularity) that would differ substantially from a black hole~\cite{Lemos2008,Danielsson2021,Chirenti2007,MichaelSMorris1988a,Simpson2019b,Visser1989a,Visser1989b,visser1989traversable1,Visinelli2021}. An object endowed with these properties is usually referred to as a \emph{black hole mimicker} (see \cite{Lemos2008} for an extensive review of such objects).

In this regard, a black hole mimicker that has gained increasing attention during the last years is the so-called black bounce,  introduced by Simpson and Visser~\cite{simpson2019black} as a family of solutions to the GR field equations. Its mathematical description is expressed in the following line element
\begin{align}\label{blackbouncemetric}
    {\rm d}s^2 = & -\left(1-\frac{2M}{\sqrt{\xi^2+\alpha^2}}\right){\rm d}t^2 + \frac{{\rm d}\xi^2}{1-\frac{2M}{\sqrt{\xi^2+\alpha^2}}} \nonumber \\
    & + (\xi^2+\alpha^2)({\rm d}\theta^2 + \sin^2\theta\, {\rm d}\varphi^2).
\end{align}
The coordinate system employed, ($t$, $\xi$, $\theta$, $\varphi$), assumes the same definition for the coordinate time $t$ and the angles $\theta$ and $\varphi$ as in the Schwarzschild coordinate system. On the other hand, the radial coordinate $\xi$, defined in the interval, $\xi\in (-\infty , +\infty)$, encodes the appearance of a non-trivial causal structure in the spacetime, depending on the value of the sole free parameter in the metric, $\alpha$, apart from the mass of the object $M$ which has the usual meaning as in the Schwarzschild spacetime. For a better understanding of the distinction between $\xi$ and the Schwarzschild radial coordinate $r$, it is useful to focus on the two-dimensional line element $\lbrace t = \text{constant}; \xi = \text{constant}\rbrace$ which yields the following.
\begin{equation}
{\rm d}s_{(2)}^2 = (\xi^2+\alpha^2)({\rm d}\theta^2 + \sin^2\theta {\rm d}\varphi^2).    
\end{equation}
The pre-factor to the standard solid angle element identifies the aeral radius of this hypersurface, given now by $r^2(\xi) = \xi^2 + \alpha^2$.
By setting $\alpha$ to zero, Eq.~\eqref{blackbouncemetric} exactly reduces to the Schwarzschild metric in Eq.~\eqref{schwmetric}, and $\xi$ to $r$. When $\alpha > 0$, on the other hand, this geometry is endowed with a \emph{throat}, coinciding with the absolute minimum of the areal radius, reached at $\xi=0$. In this case, $r_{\text{th}}^2 := r(0) = \alpha^2$ which allows for the identification of the value of $\alpha$ with the physical radius of the throat. Additionally, one can study the presence and location of the event horizon by computing the roots of the $g_{00}$ component of the metric
\begin{equation}
    1-\frac{2M}{\sqrt{\xi^2+\alpha^2}}=0 \quad \Rightarrow \quad \xi_\text{H} = \pm \sqrt{4M^2-\alpha^2}.
    \label{eq:horizons}
\end{equation}
It is clear from Eq.~\eqref{eq:horizons} that the spacetime can possess  two horizons if and only if the relation $\alpha \leqslant 2M$ is satisfied. Provided the latter inequality holds, any choice of $\alpha$ within this interval will result in a geometry with two horizons on either side of the throat that is located at $\xi = 0$. 
Consequently, the extremal value $\alpha=2M$ separates the family of black bounce solutions into types of geometries with fundamentally distinct natures:

\begin{itemize}
    \item Setting $\alpha = 0$, one obtains $\xi_\text{H} = 2M$. As mentioned, the geometry coincides with that of a standard Schwarzschild black hole of mass $M$, characterized by a real curvature singularity at the point $\xi = r = 0$.
    
    \item For $0<\alpha<2M$, the metric describes a spacetime possessing both horizons (located at $\pm\xi_\text{H
}$) and a non-zero throat (the latter being a spacelike hypersurface). However, since $r^2(\xi_\text{H}) = 4M^2$ is greater than $r^2_\text{th}$, the {\it bounce} effectively lies hidden behind the horizon. The geometry thus still represents a black hole, but the presence of the throat at $\xi = 0$ regularizes the central singularity resulting in a regular black hole~\cite{hayward2006formation, roman1983stellar, cano2019regular}. In fact, authors in Ref.~\cite{simpson2019black} showed that the non-zero components of the Riemann tensor and various curvature invariants all possess denominators with powers of ($\xi^2+\alpha^2$), which lead to the non-divergence of such curvature invariants in the limit $\xi\rightarrow 0$.
    
    \item For the critical value $\alpha = 2M$, the throat merges with the horizon and thus represents a null hypersurface~\cite{simpson2019black}. Particles can, therefore, traverse the throat from one side and then emerge at a past null horizon in the opposite region (which can be regarded as a future manifestation of the original universe), but no particle can emerge from the throat on \emph{this} side. This results in a geometry that represents a one-way traversable wormhole.
    \item Finally, for values of $\alpha > 2M$, the horizons disappear. The bounce, now a timelike hypersurface, becomes completely exposed on both sides. This configuration coincides with a two-way traversable wormhole~\cite{morris1988wormholes, visser1989traversable,visser1989traversable1,MichaelSMorris1988a},    
    with particles that can go in and out of the throat, establishing a connection between two distinct spacetime regions which can be referred to as: {\it Universe 1} (with $\xi>0$) and {\it Universe 2} (with $\xi<0$).
\end{itemize}

\subsection{Geodesics}
\label{subsec:Geodesics}

The discussion starts with the geodesic trajectories whose expressions, once derived from the line element as described in Eq.~\eqref{blackbouncemetric}, are presented in Eqs.~\eqref{Eqns_geodesics_Lagrangian_1}-~\eqref{Eqns_geodesics_Lagrangian_4}.

Within the region defined by $\alpha\leqslant 2M$, the dynamical behavior of the test particles closely mirrors that observed in Schwarzschild, see for instance Ref.~\cite{Wald:2024gr}.
Specifically, the classification of orbits in {\it plunging}, {\it bound}, and {\it unbound} trajectories is preserved. In other words, although precise geodesic trajectories have a dependence on $\alpha$ within the range $[0,\,2M]$, the fundamental categorization of these orbits remains unchanged. In contrast, in the wormhole scenario ($\alpha>2M$), this picture is drastically altered. In particular, in this case as the central object lacks an event horizon, plunging orbits are fundamentally modified. The presence of the throat, which in this case is a timelike hypersurface, allows the passage of both light and matter from both sides, requiring a finite amount of coordinate time to traverse. For this reason, null trajectories, depending on their impact parameter, either scatter off the central source, remaining confined to the original asymptotic region (for example $\xi > 0$), or plunge through the throat and propagate to infinity in the alternate Universe (with the coordinate $\xi$ evaluated along the geodesic changing sign at the throat). This represents a clear distinction from the usual black hole case, where plunging trajectories are completely absorbed at the event horizon, giving rise to the so-called \emph{black hole shadow}. 

\begin{figure*}
    \includegraphics[width = \textwidth]{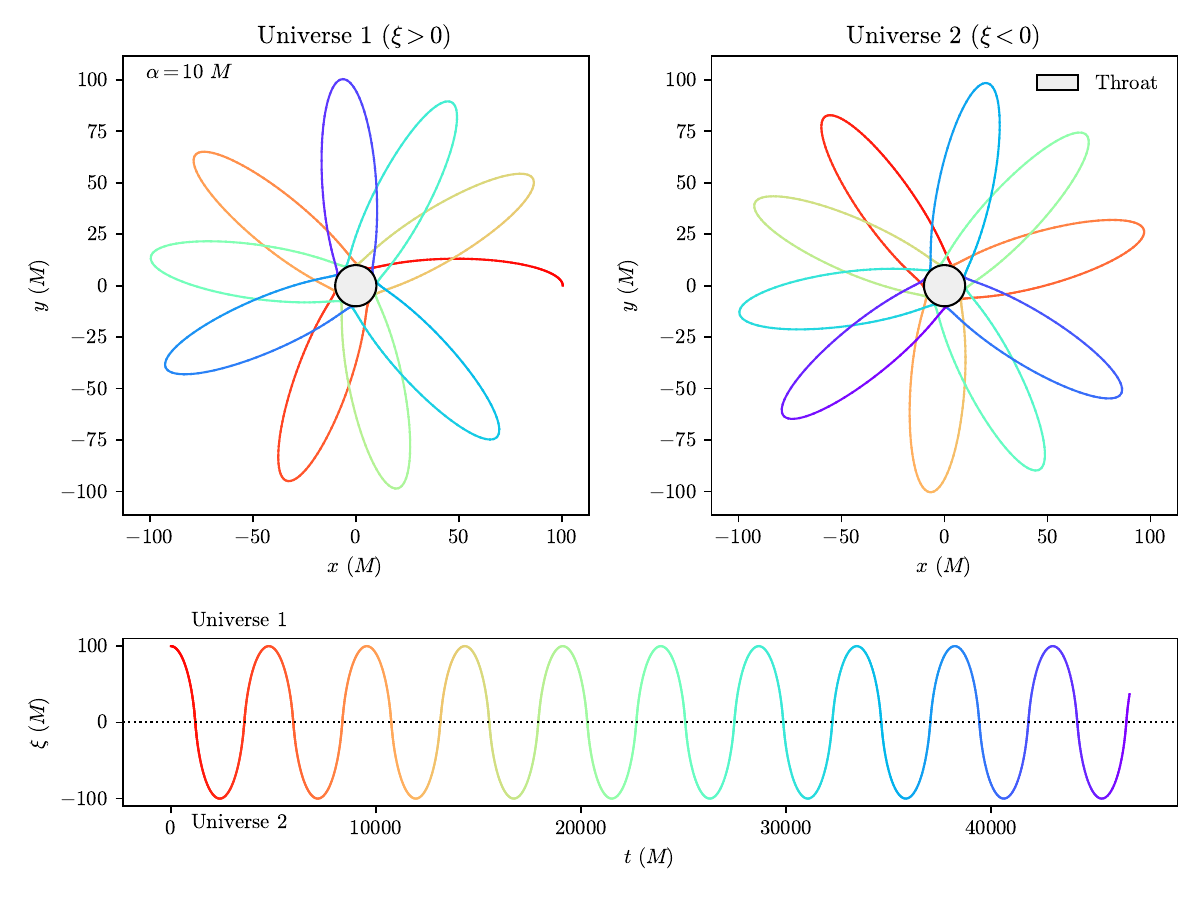}
    \caption{Time-like geodesic in a two-way traversable black bounce wormhole with $\alpha=10M$. The particle moves on the equatorial plane, starting at $\xi=100M$ with an initial tangential velocity $\simeq0.03c$. In Schwarzschild, the same initial conditions would lead to a plunging trajectory reaching the horizon only after infinite coordinate time. Here, the throat is a time-like hypersurface, allowing crossings between the two universes in finite coordinate time. The lower panel shows the coordinate $\xi$ as a function of the coordinate time $t$, which is also used as a color scale along the orbit. The trajectory undergoes periapsis precession while oscillating periodically between Universe~1 and Universe~2, completing one orbit in each universe before traversing the throat again.}
    \label{fig:timelike}
\end{figure*}

On the other hand, for timelike bound trajectories, there is no real plunging configuration in the conventional sense. Particles that have such extreme orbital parameters that their periastron distance falls \emph{inside the throat}, oscillate periodically between the two Universes. The radial coordinate thus switches sign between positive and negative values as the particle goes in and out of the throat. An example of such an orbit  is shown in Fig.~\ref{fig:timelike}.

\subsection{Black bounce spacetime in areal coordinates}

To conclude the characterization of this geometry, it is advantageous to express the line element in terms of the areal radius $r$, defined by the relation $r = \sqrt{\xi^2+\alpha^2}$. By applying the coordinate transformation ${\rm d}r = \xi {\rm d}\xi/r$, the black bounce line element in Eq.~\eqref{blackbouncemetric} can be recast as
\begin{align}
    {\rm d}s^2 = & -\left(1-\frac{2M}{r}\right){\rm d}t^2+\frac{r^3}{(r^2-\alpha^2)(r-2M)}{\rm d}r^2 \nonumber \\
    & + r^2({\rm d}\theta^2+\sin^2\theta {\rm d}\varphi^2).\label{arealblackbouncemetric}
\end{align}
In this representation, the coordinate $r$ has a precise geometric meaning: it is the \textit{areal radius}, such that any 2-sphere at a fixed $r$ possesses a surface area exactly equal to $4\pi r^2$. However, a fundamental distinction arises when comparing this with the standard Schwarzschild geometry. While in the latter $r$ can take any value in the interval (0,$\infty$), the black bounce construction imposes a strictly positive lower bound $r \geq \alpha$. The value $r = \alpha$ corresponds to $\xi = 0$, representing the ``throat'' of the geometry. Consequently, the spacetime does not contain a central point at $r = 0$, which is the primary mechanism through which the regularization of the curvature singularity is achieved.

The transition to areal coordinates offers several distinct advantages for the phenomenological analysis of this spacetime. First, it facilitates a direct comparison with the standard Schwarzschild solution. By observing Eq.~\eqref{arealblackbouncemetric}, one can immediately identify the $g_{00}$ component as identical to that of Schwarzschild, while the modification is entirely encapsulated within the radial metric component $g_{11}$. Moreover, using the same set of coordinates as in the Schwarzschild case allows us to directly compare orbits and photon trajectories that are defined from the same set of orbital parameters.
From Eq.~\eqref{arealblackbouncemetric}, the geodesic equations for a test particle take the expressions in Eqs.~\eqref{Eqns_geodesics_Lagrangian_areal_1}-~\eqref{Eqns_geodesics_Lagrangian_areal_4}, which are solved using \texttt{PyGRO} to obtain the trajectories of particles in this spacetime.

Furthermore, the areal form clarifies the nature of the horizons. The horizons remain located at $r_\text{H}=2M$, just as in the Schwarzschild case. However, the radial component now contains an additional factor $(r^2-\alpha^2)^{-1}$. While this leads to a coordinate divergence at the throat $r=\alpha$, it is essential to note that this is a coordinate artifact of the mapping $\xi\rightarrow r$. As previously discussed, the curvature invariants, such as the Kretschmann scalar $R_{\mu\nu\rho\sigma}R^{\mu\nu\rho\sigma}$, remain finite and well-behaved at the bounce point, provided $\alpha>0$.

These features of the black bounce solution make it particularly interesting, as the parameter $\alpha$ enables a smooth interpolation between completely different geometrical natures for the central object: a Schwarzschild black hole and a traversable wormhole in the Morris-Thorne sense~\cite{morris1988wormholes}. This motivates the study of potential observable quantities that could help constrain the parameter $\alpha$ and thus the intrinsic geometry of compact objects
when embedded in the black bounce spacetime. 

Although the numerical integration of the geodesic equations provides a convenient way to explore orbital dy\-na\-mics, including in the strong-field regime, our goal here is to obtain analytical insight into the deviations from the Schwarzschild solution through classical test observables. To this end, we consider the weak-field limit, where the equations simplify and allow for closed-form expressions for either the periastron precession of massive particles or the deflection of light. In order to illustrate our results, we derive the aforementioned results using two complementary formalisms: the Hamilton–Jacobi approach and a Lagrangian treatment for timelike and null trajectories, respectively.

\section{Time-like geodesics in the black bounce spacetime}
\label{sec:timelikegeo}

\subsection{Hamilton-Jacobi formalism}

For a test particle of mass $m$, motion in curved spacetime is governed by the Hamilton-Jacobi equation
\begin{equation}
    g^{\mu\nu}\frac{\partial S}{\partial x^\mu}\frac{\partial S}{\partial x^\nu} + m^2 = 0,
\end{equation}
where $g^{\mu\nu}$ is the inverse spacetime metric and $S$ corresponds to the test particle action. As is widely known, in spherically symmetric spacetimes, one can define two conserved quantities, dubbed the energy $E$ and angular momentum $L$. The conservation of these quantities allows the test particle action to be expressed in the form
\begin{equation}
    S = -Et+L\varphi+S_r(r),
\end{equation}
where $S_r(r)$ is a function of the aerial radius $r$ only. Furthermore, thanks to the rotational invariance of a spherically symmetric metric we restrict the motion to the equatorial plane $\theta = \frac{\pi}{2}$ with no loss of generality. Thus, one can recast the system of geodesic Eqs. \eqref{Eqns_geodesics_Lagrangian_areal_1}-\eqref{Eqns_geodesics_Lagrangian_areal_4} for the orbital trajectory in the black bounce spacetime to the first-order equation
\begin{widetext}
\begin{equation}\label{trajperi}
    \left(\frac{{\rm d}u}{{\rm d}\varphi}\right)^2 = \frac{E^2}{L^2} - \frac{1}{h^2} + \frac{2M}{h^2}u - \left(1 + \frac{E^2\alpha^2}{L^2} - \frac{\alpha^2}{h^2}\right)u^2 + 2M\left(1 - \frac{\alpha^2}{h^2}\right)u^3 + \alpha^2u^4 - 2M\alpha^2u^5,  
\end{equation}
\end{widetext}
where we have conveniently defined $u\equiv {r}^{-1}$ and $h\equiv L/m$.
To obtain this equation, we first substitute the action into the Hamilton-Jacobi equation using the metric form in Eq.~\eqref{arealblackbouncemetric} to derive the expression for $S_r(r)$. Subsequently, the total action is differentiated with respect to the angular momentum $L$ and, since the resulting expression is constant, we set it to zero without loss of generality. After implementing the change of variables $r = 1/u$ and differentiating $\varphi$ with respect to $u$; the final form is reached by inverting and squaring the resulting derivative.

\subsection{Analytical derivation of periastron precession }

Applying the substitution $\lambda = h^2 u/M$, to Eq. \eqref{trajperi} and expanding it to the first-order in the perturbation parameter $\sigma\equiv{M^2}/{h^2}$, where for weakly-relativistic motion $\sigma \ll 1$, we have
\begin{align}\label{perturbation}
    \left(\frac{{\rm d}\lambda}{{\rm d}\varphi}\right)^2 \,=\, & \frac{1}{\sigma}(\mathcal{E}^2-1) + 2\lambda-\left(1+\frac{\mathcal{E}^2\alpha^2\sigma}{M^2}-\frac{\alpha^2\sigma}{M^2}\right)\lambda^2 \nonumber \\
    & +2\sigma\lambda^3,
\end{align}
where $\mathcal{E} \equiv E/m$ is the energy per unit mass of the test particle. Differentiating this expression with respect to $\varphi$ yields
\begin{equation}\label{geneqmot}
    \frac{{\rm d}^2\lambda}{{\rm d}\varphi^2} + \lambda = 1+\left(\frac{\alpha^2\sigma}{M^2}-\frac{\mathcal{E}^2\alpha^2\sigma}{M^2}\right)\lambda+3\sigma\lambda^2.
\end{equation}
In the Newtonian limit ($\sigma\to 0$), the equation reduces to
\begin{equation}
    \frac{d^2\lambda_\text{N}}{d\varphi^2} + \lambda_\text{N} =1,
\end{equation}
whose solution is given by
\begin{equation}
    \lambda_\text{N} = 1 + e\cos\varphi,
\end{equation}
where we have introduced the orbital eccentricity $e$. Since the general equation of motion \eqref{geneqmot} differs from the Newtonian limit at perturbative order, we can write the general solution as
\begin{equation}
    \lambda = 1+e\cos(\varphi) + \Delta\lambda,
\end{equation}
where $\Delta\lambda$ represents a perturbation of the solution at order $\mathcal{O}(\sigma)$. By substituting this into the general relativistic equation and expanding to the first order, we obtain
\begin{align}
    \frac{{\rm d}^2\Delta\lambda}{{\rm d}\varphi^2} + \Delta\lambda \,=\, & \sigma\left(\frac{\alpha^2}{M^2}-\frac{\mathcal{E}^2\alpha^2}{M^2}+3\right)(1+e\cos(\varphi)) \nonumber \\ 
    & + 3\sigma e\cos(\varphi),
    \label{eq:perturbation_eq}
\end{align}
where the terms containing $\sigma\Delta\lambda$ have been neglected as they are of the second order in perturbations. A solution to this differential equation can be found using the ansatz
\begin{equation}
    \Delta\lambda = a_0 + a_1\varphi\sin\varphi,
\end{equation}
where $a_0$ and $a_1$ are integration constants to be determined. Substituting this test solution in Eq.~\eqref{eq:perturbation_eq}, we find that
\begin{align}
    \Delta\lambda \,=\, & \sigma\left(\frac{\alpha^2}{M^2}-\frac{\mathcal{E}^2\alpha^2}{M^2}+3\right) \nonumber \\ & + \sigma\frac{e}{2}\left(\frac{\alpha^2}{M^2}-\frac{\mathcal{E}^2\alpha^2}{M^2}+6\right)\varphi\sin(\varphi),
\end{align}
so that the full solution takes the form
\begin{align}
    \lambda = & 1 + e\cos(\varphi) + \sigma\left(\frac{\alpha^2}{M^2}-\frac{\mathcal{E}^2\alpha^2}{M^2}+3\right)\nonumber\\&+ \sigma\frac{e}{2}\left(\frac{\alpha^2}{M^2}-\frac{\mathcal{E}^2\alpha^2}{M^2}+6\right)\varphi\sin(\varphi).
\end{align}
For any positive number $\beta \ll 1$, one can resort to the following approximation
\begin{eqnarray}
    \cos[\varphi(1-\beta)] & \approx \cos(\varphi) + \beta\varphi\sin(\varphi).
\end{eqnarray}
We can apply the results above to our solution by defining $\beta \equiv \sigma\left(\frac{\alpha^2}{2M^2}-\frac{\mathcal{E}^2\alpha^2}{2M^2}+3\right)\ll1$, which allows us to finally express the general solution at first order in $\sigma$ as
\begin{widetext}
\begin{equation}
    \lambda = 1 + \sigma\left(\frac{\alpha^2}{M^2}-\frac{\mathcal{E}^2\alpha^2}{M^2}+3\right) + e\cos\left\lbrace\varphi\left[1-\sigma\left(\frac{\alpha^2}{2M^2}-\frac{\mathcal{E}^2\alpha^2}{2M^2}+3\right)\right]\right\rbrace.
    \label{lambda_final}
\end{equation}
\end{widetext}
Consequently, a full radial libration is completed over an azimuthal angle of $2\pi + \Delta\phi$, where
\begin{equation}\label{precbb}
    \Delta\phi = \frac{6\pi M}{a(1-e^2)} + \frac{\pi\alpha^2}{aM(1-e^2)}(1-\mathcal{E}^2),
\end{equation}
which we have obtained by substituting the expression defined in Eq.~\eqref{perturbation} for the perturbation parameter $\sigma$ and  expressing the quantity $h$ in terms of semi-major axis $a$ and eccentricity $e$ as $h^2 = aM(1-e^2)$~\cite{hobson2006general}.
In terms of orbital parameters, and to first order in $M/a$, this expression reads
\begin{equation}\label{precbborb}
    \Delta\phi = \frac{6\pi M}{a(1-e^2)} + \frac{\pi\alpha^2}{a^2(1-e^2)}\,.
\end{equation}
This result means that the trajectory is not a closed ellipse, leading to the precession of the periastron as per Eq. \eqref{lambda_final}.
The first term in Eq.~\eqref{precbborb} -- which is the only one surviving in the limit $\alpha\to0$ -- obviously corresponds the usual first-order expression for the orbital precession around a Schwarzschild  mass $M$. The additional term encodes the modification of the orbital trajectory arising in the black bounce spacetime. This term turns out to be quadratic in $\alpha$ implying that the larger the throat, the larger will be the departure from the usual Schwarzschild-like precession.

\subsection{Numerical periastron dynamics and associated redshift signatures}

Having derived the weak-field expression for the periastron advance in the black bounce geometry, we now compare it with direct numerical integrations of the geodesic equations performed with \texttt{PyGRO}~\cite{della2025pygro}. This comparison provides a consistency check of Eq.~\eqref{precbborb} and illustrates the effect of the bounce parameter $\alpha$ on bound timelike orbits.

\begin{figure}
\centering
\includegraphics[width=\columnwidth]{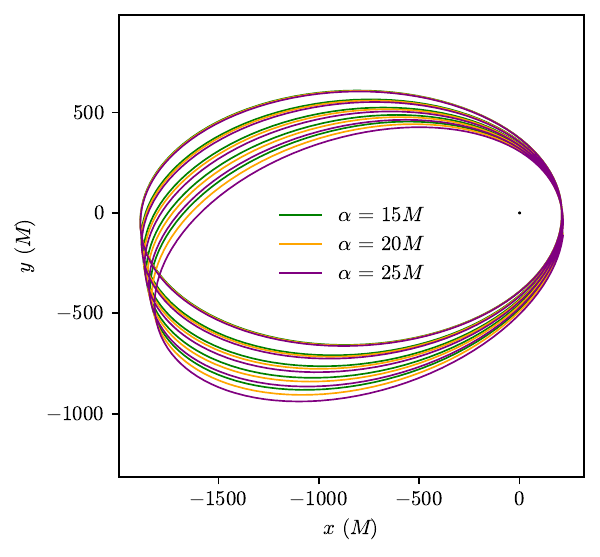}
\caption{Bound timelike geodesics in the black bounce spacetime for $a=1050M$, and $e=0.8$, with $\alpha=15M$, $20M$, and $25M$. The increase of $\alpha$ enhances the accumulated periastron advance.}
\label{precesionbbalpha}
\end{figure}

Fig.~\ref{precesionbbalpha} shows three representative trajectories with the same initial orbital parameters, $a=1050M$ and $e=0.8$, and different values of $\alpha$. Increasing $\alpha$ produces a visibly larger precession of the orbital ellipse. The orbit with $\alpha=15M$ has the smallest periastron advance, while the case $\alpha=25M$ displays the largest accumulated precession over the integration time. This behaviour is qualitatively consistent with the weak-field result, where the leading black bounce correction enters as a positive contribution proportional to $\alpha^2$ in addition to the Schwarzschild term.

\begin{figure}
\centering
\includegraphics[width=\columnwidth]{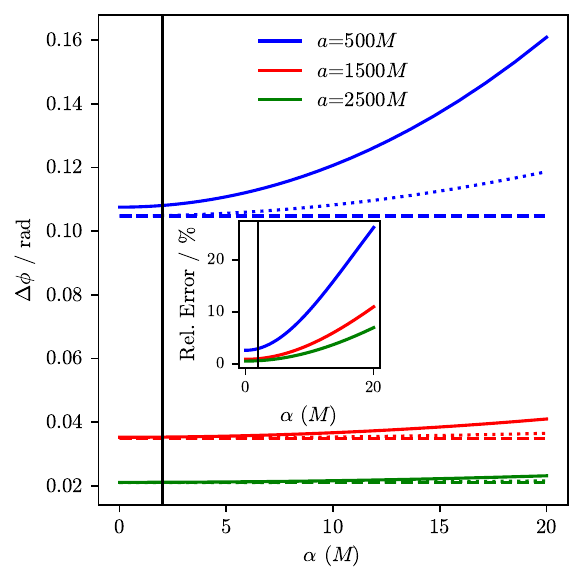}
\caption{Periastron advance as a function of the black bounce parameter $\alpha$ for $e=0.8$, considering three values of the semi-major axis $a$. Solid lines correspond to numerical geodesic integrations performed with \texttt{PyGRO}, dotted lines to the weak-field prediction of Eq.~\eqref{precbborb}, and dashed lines to the Schwarzschild weak-field result. The vertical black line marks the transition to the wormhole branch. The inset shows the relative difference between the analytical and numerical results.}

\label{figprecbb}
\end{figure}

A more quantitative comparison is shown in Fig.~\ref{figprecbb}, where the periastron advance obtained from the numerical integrations is compared with the weak-field prediction of Eq.~\eqref{precbborb}. The Schwarzschild value is also shown as a reference. For all orbital sizes considered, the black bounce precession lies above the Schwarzschild prediction and increases monotonically with $\alpha$, confirming the additive character of the leading correction. The agreement between the numerical and analytical results is best for the widest orbits, $a=1500M$ and $a=2500M$, for which the weak-field expansion is expected to be most accurate. In these cases the two curves remain close over most of the parameter range. For the more compact orbit, $a=500M$, the discrepancy is larger, especially at higher values of $\alpha$. This is not unexpected: increasing $\alpha$ strengthens the departure from Schwarzschild and, for sufficiently compact orbits, pushes the system outside the regime in which the leading weak-field approximation is quantitatively reliable. The inset of Fig.~\ref{figprecbb} shows the relative difference between the analytical and numerical results. The error remains moderate for the two widest orbits, while it increases for $a=500M$, reaching values above the percent-level accuracy expected from a leading-order weak-field treatment. Overall, the comparison confirms that Eq.~\eqref{precbborb} captures the correct dependence on $\alpha$ in the weak-field regime, while also identifying the expected loss of accuracy for smaller semi-major axes and larger bounce parameters.

\begin{figure}
\centering
\includegraphics[width=\columnwidth]{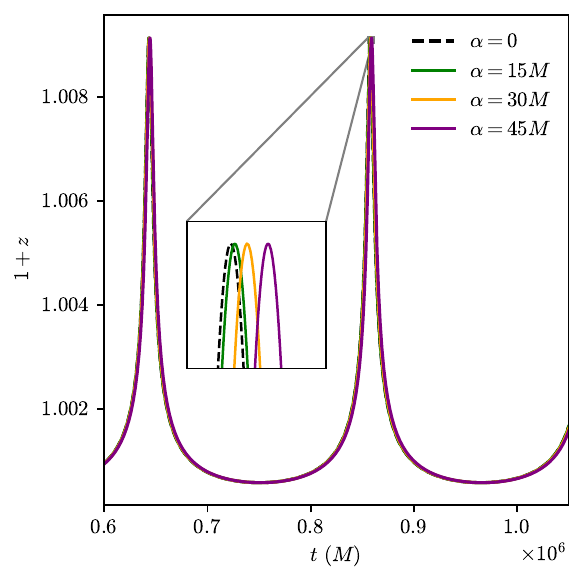}
\caption{Plot of the redshift as a function of time for $a = 1050M$ and $e = 0.8$. Solid green, yellow and purple lines re\-pre\-sent the numerical calculation performed using \texttt{PyGRO} and the black dashed line correspond to the redshift in Schwarzschild geometry.}
\label{figredshift}
\end{figure}

As an additional relativistic observable for bound time-like geodesics, we focus on the time-dilation experienced by the massive test particle, which is manifested in the form of a redshift. This is computed from the time component of the four-velocity, $1+z\equiv u^0={\rm d}t/{\rm d}\tau$, which encodes the combined effect of gravitational time dilation and the longitudinal Doppler shift for an observer at infinity. Fig.~\ref{figredshift} shows the redshift over one orbital period for the Schwarzschild limit, $\alpha=0$, and for three black bounce configurations with $\alpha=15M$, $30M$, and $45M$. The signal exhibits the expected sharp maxima at periastron, where both the gravitational potential and the orbital velocity are largest. For the orbital configurations considered here, the peak amplitude is essentially unchanged as $\alpha$ is varied, indicating that the maximum redshift at periastron is only weakly affected by the bounce parameter in this regime. Instead, the main effect of $\alpha$ is visible in the timing of the peaks. As highlighted in the inset of Fig.~\ref{figredshift}, larger values of $\alpha$ produce a cumulative phase shift in the redshift curve, reflecting the corresponding change in the radial period and in the accumulated periastron advance.

\section{Null geodesics in the black bounce spacetime}
\label{sec:nullgeo}

\subsection{Lagrangian formalism}

In the following, we shall focus our attention on the photons trajectory. In order to do so, we consider the Lagrangian $\mathcal{L} = g_{\mu\nu}\dot{x}^\mu\dot{x}^\nu$, where $\dot{x}^\mu\equiv{{\rm d}x^\mu}/{{\rm d}\tilde{\lambda}}$ and $\tilde{\lambda}$ some affine parameter. Thus, by resorting to Eq. \eqref{arealblackbouncemetric}, one can express $\mathcal{L}$ as
\begin{align}
    \mathcal{L} = &-\left(1-\frac{2M}{r}\right)\dot{t}^2 + \frac{r^3}{(r^2-\alpha^2)(r-2M)}\dot{r}^2\nonumber\\
    &+ r^2(\dot{\theta}^2+\sin^2\theta\dot{\varphi}^2).
\end{align}
The geodesic equations are obtained by substituting this expression for $\mathcal{L}$ into the Euler-Lagrange equations. In particular, we perform this by first restricting the ana\-lysis to the equatorial plane ($\theta = {\pi}/{2}$; $\dot{\theta} = 0$) without loss of generality, and second by using as the first integral of motion the normalization condition for null geodesic, i.e., $g_{\mu\nu}\dot{x}^\mu\dot{x}^\nu = 0$. In summary, the obtained equations are as follows
\begin{align}
    &\left(1-\frac{2M}{r}\right)\dot{t} = E,\label{primgeofin}\\
    &-\left(1-\frac{2M}{r}\right)\dot{t}^2 + \frac{r^3}{(r^2-\alpha^2)(r-2M)}\dot{r}^2 + r^2\dot{\varphi}^2 = 0,\label{seggeofin}\\
    &r^2\dot{\varphi} = L.\label{tergeofin}
\end{align}
An equation for the energy can be obtained by plugging Eqs. \eqref{primgeofin} and \eqref{tergeofin} into Eq. \eqref{seggeofin} and applying the fact that
\begin{equation}
    \frac{{\rm d}r}{{\rm d}\tilde{\lambda}} = \frac{L}{r^2}\frac{{\rm d}r}{{\rm d}\varphi}.
\end{equation}
Taking all this into account, one obtains
\begin{equation}\label{defbb}
    E^2 = \frac{L^2}{1-\alpha^2u^2}\left(\frac{{\rm d}u}{{\rm d}\varphi}\right)^2+L^2u^2(1-2Mu).
\end{equation}
where we have resorted again to the variable $u$ as defined immediately after Eq. \eqref{trajperi}.
Differentiating this equation with respect to $\varphi$ yields
\begin{equation}
    \frac{{\rm d}^2u}{{\rm d}\varphi^2} + u = 3Mu^2-\frac{\alpha^2}{b^2}u+2\alpha^2u^3-5M\alpha^2u^4,
    \label{Eq_u_null}
\end{equation}
where the impact parameter $b$ is defined as $b \equiv L/E$. It is worth noting that the last three terms of this equation arise exclusively from the black bounce geometry. Being strictly proportional to powers of the parameter $\alpha$, these terms encapsulate the intrinsic modifications of the null geodesic equation introduced by this metric. In the Schwarzschild limit $\alpha \rightarrow 0$, these contributions vanish, and the standard Schwarzschild orbital equation is recovered.

\subsection{Analytical derivation of light deflection}

We begin by considering Eq. \eqref{Eq_u_null}. Since we are in the weak-field regime, we neglect the term proportional to $\alpha^2u^4$, leaving the differential equation to be solved as
\begin{equation}
\label{Eq_u_complete_null_geodesics}
    \frac{{\rm d}^2u}{{\rm d}\varphi^2} + u = 3Mu^2-\frac{\alpha^2}{b^2}u+2\alpha^2u^3.
\end{equation}
The linear homogeneous version of Eq.~\eqref{Eq_u_complete_null_geodesics} possesses a general solution satisfying $u_0(\varphi=\pi/2)=1/b$ as follows
\begin{equation}
     u_0 = \frac{\sin\varphi}{b}.
\end{equation}
We will treat this solution as the zeroth-order solution for the equation of motion and write the perturbed solution of Eq. 
\eqref{Eq_u_complete_null_geodesics} as 
\begin{equation}
    u = \frac{\sin\varphi}{b}+\Delta u,
    \label{hypothesis_u_null_geodesics}
\end{equation}
where $\Delta u$ is small perturbation. Substituting the ansatz in Eq.~\eqref{hypothesis_u_null_geodesics} into Eq. 
\eqref{Eq_u_complete_null_geodesics}
and keeping terms up to first order in $\Delta u$ we obtain
\begin{equation}
     \frac{{\rm d}^2\Delta u}{{\rm d}\varphi^2} + \Delta u = \frac{3M}{b^2}\sin^2\varphi-\frac{\alpha^2}{b^3}\sin\varphi + \frac{2\alpha^2}{b^3}\sin^3\varphi,
     \label{Eq_for_Delta_u_null_geodesics}
\end{equation}
where we omit all higher-order contributions of the form $(\Delta u)^n/b^m$ for $n\geq 1$ and $m\geq 1$, consistent with our weak-field approximation. To solve the equation above, we propose the following solution
\begin{equation}
    \Delta u = a_0 + a_1\varphi\cos\varphi + a_2\cos(2\varphi) + a_3\sin(3\varphi),
\end{equation}
where $a_{0,1,2,3}$ are constants to be determined. Introducing this ansatz into Eq. \eqref{Eq_for_Delta_u_null_geodesics} and comparing the two sides of the equality, one obtains 
\begin{align}
\Delta u =  \frac{M}{2b^2}\left(3+\cos(2\varphi)\right)   
+ \frac{\alpha^2}{16b^3}\left(-4\varphi\cos\varphi+\sin(3\varphi)\right),
\end{align}
and thus the complete solution of Eq. \eqref{Eq_u_complete_null_geodesics} is
\begin{align}
u \approx & \frac{\sin\varphi}{b}+ 
\frac{M}{2b^2}\left(3+\cos(2\varphi)\right)   
+ \frac{\alpha^2}{16b^3}\left(-4\varphi\cos\varphi+\sin(3\varphi)\right).
\nonumber \\
&
\end{align}
By taking the limit $r\rightarrow\infty$ (or $u\rightarrow 0$), we analyze the small deflection case, allowing the approximations $\sin\varphi\approx\varphi$, $\cos\varphi\approx 1$, $\cos(2\varphi)\approx 1$ and $\sin(3\varphi)\approx 3\varphi$. This process results in 
\begin{equation}
    \varphi_\infty \approx -\frac{2M}{b\left(1-\frac{\alpha^2}{16b^2}\right)}.
\end{equation}
Thus, the total deflection is
\begin{equation}
    \Delta\varphi = 2|\varphi_\infty| \approx \frac{4M}{b\left(1-\frac{\alpha^2}{16b^2}\right)}.
    \label{deflection38}
\end{equation}
Since in the weak-field regime $\alpha^2\ll b^2$, we can expand the denominator in Eq. \eqref{deflection38} to first order and express the total deflection as
\begin{equation}\label{weakpred}
    \Delta\varphi = \frac{4M}{b}+\frac{M\alpha^2}{4b^3},
\end{equation}
where all terms of $\mathcal{O}(\alpha^4/b^5)$ and next are neglected.

\subsection{Numerical comparison of light deflection}

\begin{figure}[htbp]
\centering
\includegraphics[width=\columnwidth]{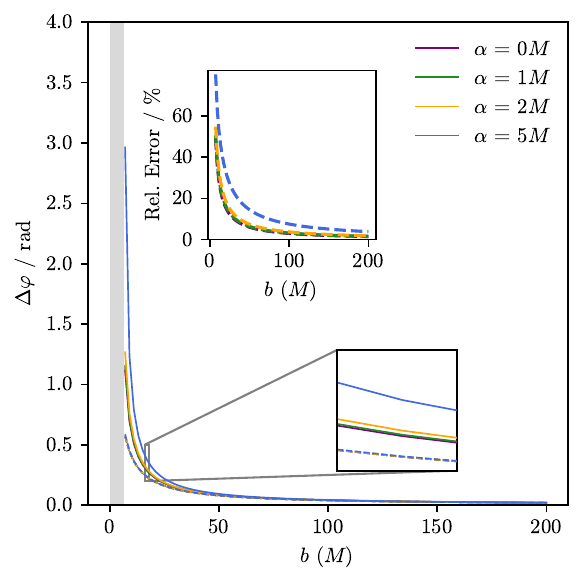}
\caption{Deflection angle versus impact parameter for photons originating at $r = 2000M$. The plot shows the deflection for black bounce parameters $\alpha = 0,\, 1M,\, 2M$ and $5M$. Solid lines repre\-sent the numerical calculation performed using \texttt{PyGRO}, while the dashed lines show the theoretical weak-field approximation derived from Eq. \eqref{defbb}. The shaded region represents the values of the impact parameter for which light deflection occurs in a second Universe for $\alpha = 5M$. An inner plot is included within the figure, which displays the percentage relative error between the numerical and theoretical curves for each $\alpha$ as a function of the impact parameter.}
\label{figdefbb}
\end{figure}

Fig.~\ref{figdefbb} shows the light-deflection angle as a function of the impact parameter for several values of the black bounce parameter $\alpha$. The numerical results obtained with \texttt{PyGRO} are compared with the weak-field prediction of Eq.~\eqref{weakpred}. For sufficiently large impact parameters, the numerical and analytical curves agree well, confirming the leading black bounce correction to the Schwarzschild deflection. The main effect of $\alpha$ is to increase the deflection angle. At fixed mass and impact parameter, larger values of $\alpha$ produce larger angular deviations, in agreement with the positive correction proportional to $\alpha^2/b^3$ in Eq.~\eqref{weakpred}. This trend is clearly visible in the weak-field region, where the photon trajectory remains far from the central object and the expansion in powers of $M/b$ and $\alpha/b$ is reliable. As expected, the agreement deteriorates at small impact parameters. In this regime the photons probe the strong-field region, and higher-order corrections not included in Eq.~\eqref{weakpred} become relevant. The discrepancy (quantified by the relative error between the numerical and analytical prediction of the deflection angle, shown in the inset plot) is especially pronounced for the $\alpha=5M$ case. This behavior indicates that the photon trajectory is no longer described by the same weak-field scattering regime and may be affected by the global structure of the black bounce geometry, including trajectories that pass through the throat.

We next determine the critical impact parameter $b_c$ for null geodesics in the black bounce spacetime. Its operational definition depends on the causal structure of the geometry. For $\alpha\leq 2M$, the spacetime contains an event horizon at
$\xi_{\rm H}=\sqrt{4M^2-\alpha^2}$, and $b_c$ separates captured photons from those scattered back to infinity. For $\alpha>2M$, the horizon is absent and the geometry describes a two-ways traversable wormhole; in this case, $b_c$ separates photons that cross the throat and reach the second asymptotic region from those that remain in the original universe and scatter back to infinity. The critical value is obtained numerically with a bisection procedure using \texttt{PyGRO}. For each value of $\alpha$, an initial interval in $b$ is chosen so as to bracket the transition between the two possible outcomes. The interval is then iteratively refined for $N=20$ steps. In the black hole regime, we impose a stopping condition for the integration of the geodesic at the horizon, and the bisection is updated according to whether the photon is captured or scattered. In the wormhole regime, the classification is instead based on the sign of the final radial coordinate: trajectories with $\xi_{\rm final}<0$ are identified as transmitted through the throat, whereas those with $\xi_{\rm final}\geq0$ are classified as scattered.

\begin{figure}[htbp]
\centering
\includegraphics[width=\columnwidth]{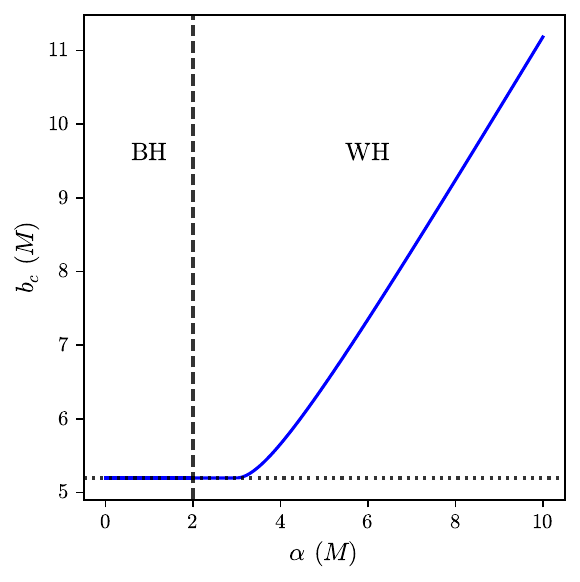}
\caption{Evolution of the critical impact parameter $b_\text{c}$ as a function of the regularization bounce parameter $\alpha$. The photons are considered to originate at a distance of $r = 2000M$. The horizontal dotted line represents the Schwarzschild limit, $b_\text{c} = 3\sqrt{3}M$. The vertical dashed line at $\alpha = 2M$ marks the topological transition between the regular black hole (BH) and the traversable wormhole (WH) regimes.}
\label{figparamimpbb}
\end{figure}

\begin{figure*}[t]
    \centering
    \subfloat{
        \includegraphics[width=0.48\textwidth]{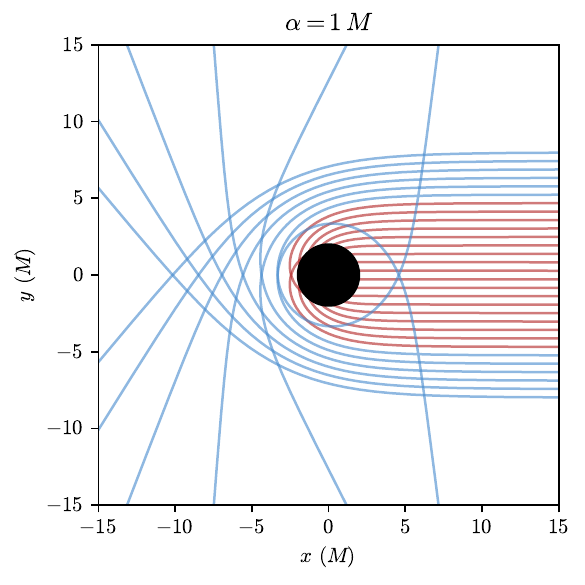}
        \label{fig:imgA}
    }
    \hfill
    \subfloat{
        \includegraphics[width=0.48\textwidth]{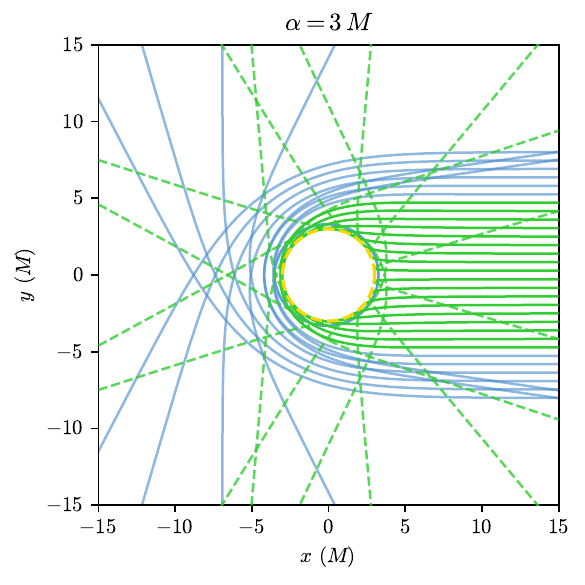}
        \label{fig:imgB}
    }
    
    \caption{Null geodesics in the black bounce spacetime for two different regimes of the regularization bounce parameter $\alpha$. \textbf{Left panel:} Regular black hole regime ($\alpha = 1M$), where the event horizon is represented by the central black disk. Red trajectories correspond to captured photons, while blue ones represent deflected rays. \textbf{Right panel:} Traversable wormhole regime ($\alpha = 3M$). The yellow dashed circle marks the throat radius. Blue trajectories are deflected within the primary universe, whereas green trajectories represent photons traversing the throat; solid lines indicate motion in the primary universe and dashed lines correspond to propagation in the secondary universe.}
    \label{Photonorbits}
\end{figure*}

We first consider the black hole branch, i.e., $\alpha\leq2M$. In this regime the numerical results, that we report in Fig.~\ref{figparamimpbb}, show that the critical impact parameter remains constant, with
$b_c\simeq5.196M$, in agreement with the Schwarzschild value
$b_c=3\sqrt{3}M$. The case $\alpha=0$ provides the expected Schwarzschild limit and serves as a useful validation of the numerical setup. More generally, the independence of $b_c$ from $\alpha$ reflects the fact that, throughout this branch, the critical photon orbit is governed by the exterior photon sphere at $r=3M$. Although the parameter $\alpha$ regularizes the central region, this modification is hidden behind the horizon and does not affect the optical capture threshold. This behavior is illustrated in the left panel of Fig.~\ref{Photonorbits}, where photon trajectories launched from infinity are shown using the areal radial coordinate for visualization. Trajectories with $b>b_c$ are scattered back to infinity, while those with $b<b_c$ cross the horizon and are captured. The transition occurs at $b=3\sqrt{3}M$, yielding the same capture cross section for all $\alpha\leq2M$. Thus, within the black hole branch, the black bounce spacetime is indistinguishable from Schwarzschild through this particular optical observable. The constancy of the critical impact parameter in the black hole branch can also be understood analytically from the standard expression for the shadow radius~\cite{psaltis2020gravitational,psaltis2008testing},
\begin{equation}
    r_{\rm sh}=\frac{r_{\rm ph}}{\sqrt{-g_{00}(r_{\rm ph})}},
\end{equation}
where $r_{\rm ph}$ denotes the photon-sphere radius. Equivalently, $r_{\rm ph}$ is determined by
\begin{equation}
    r_{\rm ph}
    =
    \sqrt{-g_{00}}
    \left(
    \frac{{\rm d}\sqrt{-g_{00}}}{{\rm d}r}
    \bigg|_{r_{\rm ph}}
    \right)^{-1}.
\end{equation}
Applying these relations to the black bounce metric in the black hole regime gives
$r_{\rm ph}=3M$ and therefore
$r_{\rm sh}=3\sqrt{3}M$. Thus the shadow radius, and equivalently the critical impact parameter, is independent of the bounce parameter $\alpha$ for $\alpha\leq2M$, in agreement with the numerical results discussed above and with previous ray-tracing studies in Ref.~\cite{Guerrero:2021ues}.

Fig.~\ref{figparamimpbb} also shows the critical impact parameter in the traversable-wormhole branch, i.e., $\alpha>2M$. Immediately above the extremal limit, $b_c$ remains close to the Schwarzschild value, $b_c=3\sqrt{3}M\simeq5.2M$, and varies only weakly with $\alpha$. This indicates that, for small throats, the threshold between reflected and transmitted null geodesics is still controlled mainly by the Schwarzschild-like exterior geometry. For larger values of $\alpha$, the curve departs from this plateau and grows monotonically, reaching $b_c\simeq11M$ at $\alpha=10M$. The increase is approximately linear over most of the range explored. We interpret this behavior in terms of the growing relevance of the presence of throat in the dynamics of null trajectories. As $\alpha$ increases, photons with larger impact parameters are transmitted to the second asymptotic region, while only those with $b>b_c$ are scattered back to the original universe. This behavior is illustrated in the second panel of Fig.~\ref{Photonorbits}, which shows representative null geodesics in the traversable-wormhole branch. Trajectories with impact parameters slightly above $b_c$ display the expected near-critical behavior, taking several turns around the throat before being scattered back to the original asymptotic region. By contrast, trajectories with $b<b_c$ cross the throat and reach the second asymptotic region, $\xi<0$. In the transformed radial coordinate used for the visualization, these transmitted trajectories appear as if they bounce at the throat radius and the portion of the trajectory that propagates on the other side of the wormhole is highlighted with a dashed line for clarity. As the throat radius grows, a larger fraction of incoming null geodesics is transmitted through the wormhole rather than scattered back. This effect provides a clear optical distinction between the traversable-wormhole branch of the black bounce spacetime and the black hole regime.

\section{Discussion and Conclusions}
\label{sec:Conclusions}

In this work, we have presented a comprehensive ana\-ly\-sis of relativistic observables in the context of the black bounce spacetime \cite{simpson2019black}, combining analytical weak-field approxi\-mations with high-precision numerical simulations. A systematic comparison with the canonical Schwarzschild geometry has revealed distinct phenomenological signatures indicative of the nonsingular nature of this class of solutions.

Our investigation of massive particle dynamics demonstrates that the regularization parameter $\alpha$ has a direct dependence on key relativistic observables like the orbital precession and the gravitational redshift. From the derived analytical expression for the periastron advance, we find that the presence of a finite throat enhances the relativistic precession. This effect is confirmed numerically, with a direct correlation between the magnitude of $\alpha$ and the precession rate. Furthermore, our analysis of the relativistic redshift profile shows that the peak redshift at periastron (where this relativistic effect is enhanced) remains essentially unchanged in the parameter range explored, while increasing $\alpha$ induces a cumulative temporal phase shift in the redshift profile, due to the modified orbital motion.

Regarding null geodesics, we found that the bounce parameter $\alpha$ systematically enhances the deflection angle of light, thereby amplifying the overall lensing effect. A particularly noteworthy result emerges from the behavior of the critical impact parameter $b_c$. In the regular black hole branch ($\alpha \leq 2M$), $b_c$ remains constant and equal to the Schwarzschild value $3\sqrt{3}\,M$, implying that the shadow radius is insensitive to the size of the internal throat as long as an event horizon is present. In contrast, once the geometry enters the traversable wormhole regime ($\alpha > 2M$), $b_c$ increases with $\alpha$ as illustrated in Fig.~\ref{figparamimpbb}, signaling a qualitative change in the effective potential. As the throat radius grows, an increasing fraction of null geodesics is transmitted to the second asymptotic region rather than scattered back. This behavior represents a clear optical signature of the throat structure and a potentially observable feature capable of distinguishing the traversable-wormhole branch from the Schwarzschild-like black hole regime.

In conclusion, this study establishes that black bounce spacetimes exhibit a rich set of observational signatures that are both qualitatively and quantitatively distinguishable from the standard Schwarzschild solution. While the black hole scenario remains largely indistinguishable from the usual case, the wormhole scenario can be effectively probed through shadow measurements and the precession of massive orbits, thereby providing robust channels for testing nonsingular gravity. Specifically, the deviation of the shadow radius and the anomalous periastron shift could be constrained using current and future data from the Event Horizon Telescope~\cite{event2023first} and the GRAVITY collaboration~\cite{abuter2020precession, DellaMonica:2021fdr}.

Future research incorporating rotation will be crucial to investigate how frame-dragging effects and shadow asymmetries distinguish these geometries from the Kerr metric~\cite{Mazza:2021rgq, Shaikh:2018kfv}. Furthermore, modeling accretion disk dynamics will allow for the prediction of specific luminosity profiles and observational signatures that could deviate from the standard black hole~\cite{Guerrero:2021ues}. These developments will be essential to further delineate the feasibility of detecting such features with next-generation astrophysical facilities, such as the next-generation Event Horizon Telescope (ngEHT) for high-resolution shadow imaging~\cite{ayzenberg2025fundamental} and the Extremely Large Telescope (ELT) for tracking stellar orbits with sub-milliarcsecond precision~\cite{sturm2024micado, rodeghiero2021performance}.

\begin{acknowledgments}
AdlCD acknowledges support from PID2024-158938NBI0 and CNS2024-154286 funded by MICIU/AEI/10.13039/501100011033  {\it ERDF A way of making Europe}; Project SA097P24 funded by Junta de Castilla y Le\'on (Spain), and NRF Grant (South Africa) CSUR23042798041.  
RDM acknowledges financial support provided by FCT – Fundação para a Ciência e a Tecnologia, I.P., through the ERC-Portugal program Project ``GravNewFields'' and also thanks the Fundação para a Ciência e Tecnologia (FCT), Portugal, for the financial support to the Center for Astrophysics and Gravitation (CENTRA/IST/ULisboa) through grant No.~\href{https://doi.org/10.54499/UID/PRR/00099/2025}{UID/PRR/00099/2025} and grant No.~\href{https://doi.org/10.54499/UID/00099/2025}{UID/00099/2025}.
\end{acknowledgments}

\appendix 

\section{Geodesic equations in the black bounce spacetime}
In this appendix, we present the explicit form of the geodesic equations for the black bounce spacetime, which are required for the numerical integrations carried out in this work. The equations of motion are obtained from the Lagrangian $\mathcal{L} = \frac{1}{2}g_{\mu\nu}\dot{x}^\mu\dot{x}^\nu$, where the dot denotes differentiation with respect to the proper time $\tau$ for massive particles, or with respect to the affine parameter $\tilde{\lambda}$ for null geodesics.

{\bf Areal coordinates:}
By performing the transformation $r = \sqrt{\xi^2+\alpha^2
}$, we express the metric \eqref{blackbouncemetric} in the areal representation shown in Eq.~\eqref{arealblackbouncemetric}. The resulting geodesic equations for the coordinates ($t, r, \theta, \varphi$) are
\begin{align}
    \ddot{t} = & \frac{2 M \dot{r} \dot{t}}{r\left(2 M - r\right)}\,, \label{Eqns_geodesics_Lagrangian_areal_1} \\
    \ddot{r} = & \frac{\Psi}{r^5}(\dot{\varphi}^2r^3\sin^2\theta+\dot{\theta}^2r^3-M\dot{t}^2)\, \nonumber \\ 
    & - \frac{\dot{r}^2}{2r^3}\left[\frac{r^3(4Mr+\alpha^2-3r^2)}{\Psi}+3r^2\right]\,, \label{Eqns_geodesics_Lagrangian_areal_2}\\
    \ddot{\theta} = & - \frac{2 \dot{r} \dot{\theta}}{r} + \frac{\dot{\varphi}^{2} \sin{\left(2\theta \right)}}{2
    }\,, \label{Eqns_geodesics_Lagrangian_areal_3}\\
    \ddot{\varphi} = & - \frac{2 \dot{\varphi} \dot{r}}{r} - 2 \dot{\varphi} \dot{\theta}\cot\theta \label{Eqns_geodesics_Lagrangian_areal_4}\,,
\end{align}
where
\begin{equation}
    \Psi = \Psi(r) \equiv 2M\alpha^2-2Mr^2-\alpha^2r+r^3\,.
\end{equation}

{\bf Simpson-Visser coordinates}: In the original Simpson-Visser coordinate system ($t, \xi, \theta, \varphi$), the geodesic equations are given by

\begin{eqnarray}
    \ddot{t} & =& -\frac{2M\xi\dot{t}\dot{\xi}}{\Xi(\alpha^2+\xi^2)^{3/2}}, 
    \label{Eqns_geodesics_Lagrangian_1}\\ 
    \ddot{\xi} & =& \frac{M\xi(\dot{\xi}^2-\dot{t}^2\Xi^2)}{\Xi(\alpha^2+\xi^2)^{3/2}}-\xi\Xi( \dot{\varphi}^2\sin^2\theta+\dot{\theta}^2),\label{Eqns_geodesics_Lagrangian_2} 
    \\ 
    \ddot{\theta} & =& -\frac{2\xi\dot{\xi}\dot{\theta}}{\alpha^2+\xi^2} + \frac{\dot{\varphi}^2\sin(2\theta)}{2}, 
    \label{Eqns_geodesics_Lagrangian_3} \\ 
    \ddot{\varphi} & =& - \frac{2\xi\dot{\varphi}\dot{\xi}}{\alpha^2+\xi^2}-2\dot{\varphi}\dot{\theta}\cot{\theta},
    \label{Eqns_geodesics_Lagrangian_4}
\end{eqnarray}
where
\begin{equation}
    \Xi=\Xi(\xi) \equiv 1-\frac{2M}{\sqrt{\xi^2+\alpha^2}}\,.
\end{equation}

\bibliographystyle{apsrev4-1}
\bibliography{References}

\end{document}